# Generating maximally entangled distant pair in invariant stratification spin networks

M. Ghojavand [1], M. A. Jafarizadeh [2, 3] and S. Rouhani [1]

1 Physics Department, Sharif University of Technology, P.O. Box 11365-9161, Tehran, Iran
2 Department of Theoretical Physics and Astrophysics, University of Tabriz, Tabriz 51664, Iran
3 Research Institute for Fundamental Sciences, Tabriz 51664, Iran

ghojavand@gmail.com, jafarizadeh@tabrizu.ac.ir, srouhani@sharif.ir

**Abstract**

In this paper we study the generation of Bell states between distant vertices in a permanently coupled quantum spin network, interacting via *invariant stratification graphs*. To begin with we establish a class of upper bounds over achievable entanglement between the reference site and various vertices. We observe that the maximum of these upper bounds is 1 e-bit. We conclude that the reference site can generate a Bell state with a vertex if the corresponding upper bound of the vertex is 1 e-bit. Thus for generation of a Bell state this upper bound must be saturated. Taking this into account, we obtain the characteristic constraint of the proper graphs. We introduce a special class of antipodal invariant stratification graphs, which is called *reflective,* whereas the antipode vertex obeys the characteristic constraint. We also show that the antipodal *association scheme graphs* are reflective so Bell states can be generated between the antipodal vertices. Moreover we observe that in such graphs the proper Hamiltonian that enables creation of Bell state is the Heisenberg interaction between vertex pairs.

## 1. Introduction

Entanglement is a basic ingredient in many quantum information tasks, such as precise transfer of quantum states [1], parallel processing [2] [3], and dense coding[4]. In particular maximally entangled pairs or Bell states are used in such protocols. So it is desirable to generate such pairs between different qubits. Various physical systems can act as quantum channels for generating entanglement between different qubits, one of them being a quantum spin system. In these systems qubits interact through a suitable Hamiltonian, e.g. Heisenberg or XY-Hamiltonian. In such systems two scenarios may be envisaged for entanglement of two distant qubits. Firstly, there is no interaction with the rest of system and interaction is exerted by an externally operating gate. By multiple application of controlled swap operations along the communication line entanglement occurs between the targeted qubits. However each external manipulation induces accumulating noises in the system, and may not be desirable. The essence of using more constituents for in-situ control of the quantum dynamics also limits miniaturization of the system. This makes nano-scale quantum information processors extremely difficult to approach. In the second scenario we use some qubits that naturally interact with each other. Due to these inherent interactions entanglement may be generated through the natural evolution of quantum state of the system. The desirable spin system would be one in which the initial state is simple to prepare and the generated entanglement is maximal. Our proposal for such a system is to arrange a spin system on a network given by an interaction map using graph theory. In the previous work we studied the optimization of entanglement between qubits located in the same strata in a special type of *invariant stratification graphs* (I.S.G.), which are "*underlying networks of group association schemes*" [5]. In the present work we study the problem of optimization of entanglement between reference site and the other vertices in I.S.G's. Here we show that if the qubits interact via proper I.S.G's then a simple disentangled initial quantum state evolves such that a Bell state is generated between reference site and a distant qubit.

The organization of this paper is as follows: in section 2 we recall some preliminary material on graphs and in particular I.S.G's. In section 3 we study general conditions that lead to generation of Bell state between distant vertices in I.S.G's. In sections 4 we introduce some antipodal I.S.G's that their antipodes hold the general conditions obtained in section 3. In section 5 we present some examples of the graphs introduced in section 4. The paper is closed with a summary in section 6.

## 2. Preliminaries

### Description of the system

Consider $n$ interacting qubits. Our problem is to propose proper interactions to generate a highly entangled qubit pair from a simple disentangled initial state. Here we employ graph concepts to devise such system and search for proper graphs. The Hilbert space of these qubits is $(\mathbb{C}^2)^{\otimes n}$. We confine our study to total angular momentum preserving Hamiltonians, so $\mathbf{H}$ is block diagonal $\mathbf{H} = \mathbf{H}^{(0)} \oplus \mathbf{H}^{(1)} \oplus \mathbf{H}^{(2)} \oplus ... \oplus \mathbf{H}^{(n)}$. Moreover we assume that our initial state is a simple one-particle state $|o\rangle$ and we call $o$ the reference site. The one-particle subspace is defined as $K = span\{|k\rangle : 1 \leq k \leq n\}$ where $|k\rangle$ is a state in which the spin of $k$-th qubit is upward and the others are downward. Such Hamiltonian restricts the evolved state to the one particle subspace $K$. So the quantum dynamics is determined by $\mathbf{H}^{(1)}$ which is an $n \times n$ Hermitian matrix. Hereafter we attempt to find proper $n \times n$ Hamiltonians by employing graph concepts. In this paper we use the algebraic properties of *I.S.G.* to find proper Hamiltonians as functions of graph adjacency matrix $\mathbf{A}$, thus we assume $\mathbf{H} = f(\mathbf{A})$ where $f$ is a polynomial.

### Graphs and their stratifications

A graph is a pair $\Gamma = (V, E)$ where $V$ is a non-empty set called vertex set and $E$ is a symmetric subset of $\{(x, y) : x, y \in V\}$ called the edge set of the graph. Two vertices $x, y$ are called *adjacent* if $(x, y) \in E$ and in that case we write $x \sim y$. For a graph $\Gamma = (V, E)$ the *adjacency matrix* **A** is defined as

$$(\mathbf{A})_{x,y} = \begin{cases} 1 & \text{if } x \sim y \\ 0 & \text{otherwise} \end{cases} \tag{2-1}$$

The *degree* or *valency* of a vertex, $x$ is defined as $\kappa(x) = |\{z \in V : z \sim x\}|$ identifying the number of its adjacent vertices. A finite sequence $x_0, x_1, ..., x_l \in V$ is called a walk of length $l$ (or $l$ steps) if $x_{i-1} \sim x_i$ for all $i = 1, 2, ..., l$. For $x \neq y$ let $\partial(x, y)$ be the length of the shortest walk connecting $x$ and $y$. We define *i*-th *adjacency matrix* $\mathbf{A}_i$ as

$$(\mathbf{A}_i)_{x,y} = \begin{cases} 1 & \partial(x, y) = i \\ 0 & \text{otherwise} \end{cases} \tag{2-2}$$

Now we fix a vertex $o$ as the origin of the graph. Then the graph is stratified into a disjoint union of strata by the classification $V_i(o) = \{y \in V : \partial(o, y) = i\}$ such that

$$V = \bigcup_{i=0}^{d} V_i(o) \tag{2-3}$$

where $d$ denotes the furthest strata from the reference site $o$ and is called *diameter of the graph*. Now we associate a unit vector $|\phi_i\rangle := \frac{1}{\sqrt{n_i}} \sum_{k \in V_i} |k\rangle$ to each strata and call it the *i*-th *stratum state*, with $n_i = |V_i(o)|$, the number of elements of the *i*-th stratum. Noting that $|\phi_0\rangle = |o\rangle$, we have

$$\mathbf{A}_i |\phi_0\rangle = \sqrt{n_i} |\phi_i\rangle \tag{2-4}$$

whereas the reference site can be chosen arbitrarily. The set of unit vectors $\{|\phi_i\rangle\}$ defines a subspace of one-particle states as $K' = span\{|\phi_i\rangle : 0 \leq i \leq d\}$ and is called the *stratification subspace*. We shall consider spin systems that the evolution of $|\phi_0\rangle$ -as initial state- is restricted to remain in this subspace. The physical interest for considering such systems is that the probability of finding excitation on equal strata vertices is equal at any time. For more details about these graphs see [5][6] [7]

### Invariant stratification spin networks
A subspace is *invariant* w.r.t. an operator if this subspace is closed under the action of that operator. Accordingly, we consider the graphs that one of their stratification subspaces is invariant w.r.t. the adjacency matrix. We call this type of graphs as "*invariant stratification graphs*" that is also called as *QD graphs* in [8-11]. It is easy to see that the output of the action of adjacency matrix on the one-particle states, belonging to the $i$-th stratum, is a superposition of one-particle states, belonging to $(i-1)$, $i$, and $(i+1)$ strata [8][10]. So for I.S.G's we have

$$\mathbf{A}|\phi_i\rangle = \sqrt{\omega_{i+1}}|\phi_{i+1}\rangle + \alpha_i|\phi_i\rangle + \sqrt{\omega_i}|\phi_{i-1}\rangle \tag{2-5}$$

where the set of $\{\omega_i\}$ and $\{\alpha_i\}$ are called Szego-Jacobi coefficients of the graph. The following theorem determines the necessary and sufficient condition for a graph to be an I.S.G.

Theorem 1: Let $\Gamma = (V, E)$ be a graph with a fixed origin $o \in V$. According to stratification around this vertex, let $x \in V_l$, we set

$$\kappa_\delta(x) := |\{z \in V_{l+\delta} : z \sim x\}|, \qquad \delta \in \{-1, 0, +1\}.$$

Then $\Gamma$ is an I.S.G. w.r.t. $o$ if and only if $\kappa_\delta(x)$ depends only on $l$ (does not depend on $x \in V_l$). For such graphs the Szego-Jacobi coefficients are $\omega_l = \frac{n_l}{n_{l-1}} \kappa_{-1}^2(x)$ and $\alpha_l = \kappa_0(x)$.

The above theorem says that the necessary and sufficient condition for a graph to be an I.S.G. is that the number of outgoing edges from each vertex of stratum $i$ to stratum $i-1$ (and also distinctly to strata $i$ and $i+1$) to be equal for all vertices of stratum $i$, for all $i$ [7].

Theorem 2: the strata in *antipodal I.S.G.'s* under the change of reference site to antipode vertex remains invariant and the order of Szego-Jacobi parameters is inverted as

$$|\phi_i'\rangle = |\phi_{d-i}\rangle, \qquad n_i' = n_{d-i} \tag{2-6}$$

$$\omega_i' = \omega_{d-i}, \qquad \alpha_i' = \alpha_{d-i} \tag{2-7}$$

where the prime denotes stratification around the new vertex.

Proof: due to the theorem (1), in an I.S.G each vertex located in $i$-th stratum is connected directly to $i-1$ and $i+1$ strata. This causes the strata to remain invariant under the change of reference site to antipode vertex and we achieve (2-6). Inserting (2-6) in (2-5), the relevant Szego-Jacobi parameters read as (2-7).

From the three term recursive relation (2-5) it is easily seen that

$$|\phi_i\rangle = P_i(\mathbf{A})|\phi_0\rangle \tag{2-8}$$

where the polynomial $P_i(\mathbf{A})$ is of degree i and is defined recurrently due to (2-5) as

$$\widetilde{P}_0(x) = 1, \quad \widetilde{P}_1(x) = x - \alpha_0$$

$$x\widetilde{P}_i(x) = \widetilde{P}_{i+1}(x) + \alpha_i \widetilde{P}_i(x) + \omega_i \widetilde{P}_{i-1}(x), \qquad 1 \leq i \leq d \qquad (2\text{-}9)$$

where we have defined a rescaled function as $\widetilde{P}_i(x) := \sqrt{\omega_1 \omega_2 \ldots \omega_i} P_i(x)$.

We aim to calculate the amplitudes

$$\gamma_i = \langle \phi_i | U(t) | \phi_0 \rangle = \langle \phi_i | e^{-if(\mathbf{A})t} | \phi_0 \rangle \qquad (2\text{-}10)$$

finally to optimize them, we have:

$$\gamma_i = \langle \phi_0 | \sum_{l,k} P_i(a_l) e^{-if(a_l)t} | a_l, k \rangle \langle a_l, k | \phi_0 \rangle \qquad (2\text{-}11)$$

where $a_l$'s are eigenvalues of $\mathbf{A}$, and $k$ labels the degeneracy of $a_l$. We define $\mathbf{E}_l := \sum_k |a_l, k\rangle\langle a_l, k|$ and accordingly the coefficients $w_l := \langle \phi_0 | \mathbf{E}_l | \phi_0 \rangle$, where $\mathbf{E}_l$ projects each vector onto the span of eigenvectors that correspond to $a_l$. Accordingly (2-11) reforms as

$$\gamma_i = \sum_{l=0}^{d} P_i(a_l) e^{-if(a_l)t} w_l \qquad (2\text{-}12)$$

Note that for identification of amplitudes $\gamma_i$ it is sufficient to identify $w_l$. The above calculation does not limit to identification of expectation value of the state $|\phi_0\rangle$ of operator $P_i(\mathbf{A}) e^{-if(\mathbf{A})t}$ and we can indeed identify expectation value of any function of $\mathbf{A}$ if $w_l$ is determined. In fact by exactly the same procedure we can show

$$\langle \phi_0 | g(\mathbf{A}) | \phi_0 \rangle = \sum_l g(a_l) w_l \qquad (2\text{-}13)$$

where $g(x)$ is an arbitrary function. The method of identification of expectation value of an arbitrary function of an operator through identification of its coefficients $w_l$ is called *spectral distribution technique*. As an important consequence of (2-13) we obtain some useful orthogonality relations over different polynomials $P_i(x)$ and $P_j(x)$ through a proper measure,

$$\delta_j^i = \langle \phi_j | \phi_i \rangle = \langle \phi_0 | P_j(\mathbf{A}) P_i(\mathbf{A}) | \phi_0 \rangle = \sum_{l=0}^{d} P_i(a_l) P_j(a_l) w_l \qquad (2\text{-}14)$$

Introducing two matrices $\mathbf{P}_{ij} := P_i(a_j)$ and $\mathbf{W} := diag(w_1, w_2, \ldots, w_d)$ we rewrite (2-14) in a matrix form

$$\mathbf{PWP}^t = \mathbf{1} \quad \rightarrow \quad \mathbf{PW} = (\mathbf{P}^t)^{-1} \qquad (2\text{-}15)$$

Fortunately the spectral distribution of symmetric tridiagonal matrices of the type (2-5) can be identified through a standard procedure. The eigenvalues of these matrices are the roots of

$$\widetilde{P}_{d+1}(x) = 0 \tag{2-16}$$

Introducing a new polynomial $Q_d(x)$ through the following three-term recursion relation

$$Q_0(x) = 1, \quad Q_1(x) = x - \alpha_1$$

$$xQ_i(x) = Q_{i+1}(x) + \alpha_{i+1}Q_i(x) + \omega_{i+1}Q_{i-1}(x) \tag{2-17}$$

the corresponding quadrature weight $w_l$ is obtained as

$$w_l = \lim_{x \to a_l} \frac{(x - a_l)Q_d(x)}{\widetilde{P}_{d+1}(x)} \tag{2-18}$$

For more information about I.S.G. see [5][6][7][9]

## 3. Entanglement and its optimization

In a one-particle state $|\psi\rangle = \sum_k \upsilon_k |k\rangle$ it can be shown that the entanglement between qubits $m$ and $n$ is $C_{mn}(t) = 2|\upsilon_m(t)\upsilon_n(t)|$ [5] . So here the entanglement between two qubits located in two strata $0$ and $i$ is

$$C_{0i}(t) = \frac{2|\gamma_0(t)\gamma_i(t)|}{\sqrt{n_i}} \tag{3-1}$$

The probability conservation rule $\sum_{k=0}^{d}|\gamma_k(t)|^2 = 1$ establishes an upper bound on the attainable entanglement. Consider the ultimate situation where excitations are completely localized on the targeted strata $0$ and $i$ so that, $|\gamma_0|^2 + |\gamma_i|^2 = 1$ . Assuming this constraint, the maximum of $|\gamma_0\gamma_i|$ is attained when

$$|\gamma_0| = |\gamma_i| = \frac{1}{\sqrt{2}} \tag{3-2}$$

thus the entanglement upper bound is found to be

$$C_{0i\_UB} = \frac{1}{\sqrt{n_i}} \tag{3-3}$$

According to (3-3) $n_i$ must be as low as possible to achieve a rather large amount of entanglement between reference site and the elements of $i$-th stratum. In particular if $n_i = 1$

(e.g. antipodal I.S.G's) we hope to generate a maximally entangled vertex pair. In the following subsection we investigate the conditions for saturating the above upper bound.

## Saturating entanglement upper bounds

Here we study the possibility of saturating the upper bound (3-3) by using a suitable Hamiltonian $\mathbf{H} = f(\mathbf{A})$. By introducing a new variable

$$\eta_l(t) := e^{-if(a_l)t} \tag{3-4}$$

if we rewrite the set of equations (2-12) in a vector form, $\mathbf{PW}\vec{\eta} = \vec{\gamma}$, and using (2-15) we find

$$\vec{\eta} = \mathbf{P}^t \vec{\gamma} \tag{3-5}$$

Considering the (3-2) case that is $\vec{\gamma} = \frac{1}{\sqrt{2}}\left(e^{i\xi_0}|\phi_0\rangle + e^{i\xi_i}|\phi_i\rangle\right)$ we obtain

$$\eta_l = \frac{1}{\sqrt{2}}(P_{0l}e^{i\xi_0} + P_{il}e^{i\xi_i}) = \frac{e^{i\xi_0}}{\sqrt{2}}(1 + P_{il}e^{i\xi_\Delta}) \tag{3-6}$$

on the other hand, from (3-4) the constraints $|\eta_l| = 1$ must be satisfied for all l, so

$$\frac{1}{2} + \frac{P_{il}^2}{2} + P_{il}Cos\xi_\Delta = 1 \tag{3-7}$$

as equation (3-7) holds for all of the elements of $i$-th row of $\mathbf{P}$, $P_{il}$'s, are necessarily one of the roots of the following quadratic equation

$$Y^2 + (2Cos\xi_\Delta)Y - 1 = 0 \tag{3-8}$$

This equation always has two distinct real roots. This means that the $i$-th row of $\mathbf{P(A)}$ must be composed from at most two distinct values $p_i^+$ and $p_i^-$ such that

$$p_i^+ p_i^- = -1 \tag{3-9}$$

Only for such graphs we can find the proper $\mathbf{H} = f(\mathbf{A})$ by taking part the other equality

$$p_i^+ + p_i^- = -2Cos\xi_\Delta \tag{3-10}$$

Inserting (3-4) in (3-6) and considering (3-10) we achieve

$$f(a_l)t^* = 2n_l\pi + \xi_0 + tg^{-1}\left(\frac{P_{il}Sin\xi_\Delta}{1 + P_{il}Cos\xi_\Delta}\right) = 2n_l\pi + \xi_0 \pm tg^{-1}\left(\frac{\sqrt{-P_{il}^4 + 6P_{il}^2 - 1}}{3 - P_{il}^2}\right) \tag{3-11}$$

where $n_l$ are arbitrary integers and $t^*$ is a fixed time for which (3-9) is met. Now we can determine a polynomial $f$ of degree $d$ that satisfies (3-11). If we define $\tau_l := f(a_l)t^*$, we have

$$f(x)t^* = \sum_{l=0}^{d} J_l P_l(x) \quad \rightarrow \quad \tau_l = \sum_{k=0}^{d} J_k P_{kl} \quad \rightarrow \quad \vec{\tau} = \mathbf{P}^t \vec{J} \tag{3-12}$$

using (2-15) the coefficients $\vec{J}$ are obtained as

$$\vec{J} = \mathbf{PW}\vec{\tau} \tag{3-13}$$

In next sections we introduce some classes of antipodal I.S.G's that their antipodes become maximally entangled.

## 4. Bell state generation between antipodes in invariant stratification graphs

### Antipodal mirror symmetric I.S.G's

We call an antipodal I.S.G *reflective* if the sequence of Szego-Jacobi parameters remains invariant under the change of reference site to antipode vertex

$$\omega'_i = \omega_i, \qquad \alpha'_i = \alpha_i \tag{4-1}$$

For instance (4-1) holds for every *I.S.G* that (their vertices and edges) remain invariant under the reflection w.r.t. a symmetry line which is parallel to strata, (see Fig 1)

So due to (2-7) in such graphs

$$\omega_i = \omega_{d-i}, \qquad \alpha_i = \alpha_{d-i}, \qquad n_i = n_{d-i} \tag{4-2}$$

As a result of (4-1) due to (2-5)

$$P'_i(x) = P_i(x) \tag{4-3}$$

Combining (2-8), (2-6) and (4-3) we have $|\phi_d\rangle = P_d(\mathbf{A})|\phi_0\rangle$ and $|\phi_0\rangle = P_d(\mathbf{A})|\phi_d\rangle$. Consequently we obtain

$$P_d(\mathbf{A})^2 |\phi_0\rangle = |\phi_0\rangle \tag{4-4}$$

Considering (2-8) we have

$$P_d(\mathbf{A})^2 |\phi_i\rangle = P_d(\mathbf{A})^2 P_i(\mathbf{A})|\phi_0\rangle = P_i(\mathbf{A})P_d(\mathbf{A})^2 |\phi_0\rangle = P_i(\mathbf{A})|\phi_0\rangle = |\phi_i\rangle \tag{4-5}$$

Since (4-5) holds for all $i$ we have

$$P_d(\mathbf{A})^2 = \mathbf{1}^{(d+1)} \qquad (4\text{-}6)$$

So $P_d(a_k)^2 = 1$[1] and $a_k$ are eigenvalues of $\mathbf{A}$. This means that $P_{dk}$ are +1 or -1 thus the constraint (3-9) is satisfied. So we can generate maximal entanglement (1 e-bit) between the reference site and its antipode which is the farthest vertex from it. In the next subsection we introduce an important class of reflective antipodal I.S.G's and study its proper Hamiltonian in details.

### Association scheme graphs

For each two vertices, x and y, we define a set of numbers $\chi(x,y;i,j) := |\{z \in V : \partial(x,z) = i \ \& \ \partial(z,y) = j\}|$. If $\chi(x,y;i,j) = g(\partial(x,y);i,j)$ then we call the graph as an *association scheme* and term $p_{i,j}^k := g(k;i,j)$ as *intersection numbers* of the graph. From the definition of association schemes we can easily check that $p_{i,i}^0 = n_i$.

It can be shown that in association schemes the set of adjacency matrices obey the Bose-Mesner algebra

$$\mathbf{A}_i \mathbf{A}_j = \sum_{k=0}^{d} p_{i,j}^k \mathbf{A}_k \qquad (4\text{-}7)$$

Considering an arbitrary vertex x as the reference site, $|\phi_0\rangle = |x\rangle$, we have

$$\mathbf{A}|\phi_i\rangle = \frac{1}{\sqrt{p_{i,i}^0}} \mathbf{A}\mathbf{A}_i|\phi_0\rangle = \sum_{k=0}^{d} \frac{p_{1,i}^k}{\sqrt{p_{i,i}^0}} \mathbf{A}_k|\phi_0\rangle = \sum_{k=0}^{d} \sqrt{\frac{p_{k,k}^0}{p_{i,i}^0}} p_{1,i}^k |\phi_k\rangle \qquad (4\text{-}8)$$

This shows that $\mathbf{A}|\phi_i\rangle \in \mathrm{span}\{|\phi_k\rangle : 0 \le k \le d\}$ that means every association scheme graph is an *I.S.G*. Comparing (4-8) with (2-5) we conclude that only $p_{1,i}^{i+1}$, $p_{1,i}^i$, and $p_{1,i}^{i+1}$ among $p_{1,i}^k$ may be nonzero integers, thus

$$\mathbf{A}|\phi_i\rangle = \sqrt{\frac{p_{i+1,i+1}^0}{p_{i,i}^0}} p_{1,i}^{i+1} |\phi_{i+1}\rangle + \sqrt{\frac{p_{i,i}^0}{p_{i,i}^0}} p_{1,i}^i |\phi_i\rangle + \sqrt{\frac{p_{i-1,i-1}^0}{p_{i,i}^0}} p_{1,i}^{i-1} |\phi_{i-1}\rangle \qquad (4\text{-}9)$$

and the proper Szego-Jacobi coefficient read as

$$\omega_i = \frac{p_{i-1,i-1}^0}{p_{i,i}^0} p_{1,i}^{i-1^2}, \qquad \alpha_i = p_{1,i}^i \qquad (4\text{-}10)$$

Since the reference site x is chosen arbitrarily in (4-8), we conclude in association scheme graphs all of stratification subspaces are invariant under the action of the adjacency matrix, and their corresponding Szego-Jacobi coefficients are the same and given by (4-10). So under the change of reference site the Szego-Jacobi sequence remains invariant

$$\omega'_i = \omega_i, \qquad \alpha'_i = \alpha_i \tag{4-11}$$

which shows that *antipodal association scheme graphs* are *reflective antipodal I.S.G's*. So as we mentioned earlier we can generate maximal entanglement (1 e-bit) between the antipodal vertices in such graphs.

---

[1] it can be easily shown in the diagonal representation of **A**

Considering two arbitrary vertices x and y and assuming x as the reference vertex, by using (2-4) and (2-8) we have

$$\langle x | \mathbf{A}_i | y \rangle = \langle \phi_0 | \mathbf{A}_i | y \rangle = \sqrt{n_i} \langle \phi_i | y \rangle = \langle \phi_0 | P_i(\mathbf{A}) | y \rangle = \langle x | P_i(\mathbf{A}) | y \rangle \tag{4-12}$$

Since the above relation holds for all vertex pairs x and y we conclude that

$$\mathbf{A}_i = P_i(\mathbf{A}) \tag{4-13}$$

Consequently

$$\mathbf{H} = \frac{1}{t^*} \sum_{i=0}^{d} J_i P_i(\mathbf{A}) = \frac{1}{t^*} \sum_{i=0}^{d} J_i \mathbf{A}_i \tag{4-14}$$

In the following we show that the adjacency matrices $\mathbf{A}_i$ of association schemes are equivalent to Heisenberg interactions. Consider an arbitrary vertex x as the reference site

$$\frac{1}{2} \sum_{k,l: \partial(k,l)=i} \vec{\sigma}_k . \vec{\sigma}_l | x \rangle = \sum_{k,l: \partial(k,l)=i} (\pi_{kl} - \frac{1}{2}\mathbf{1}) | \phi_0 \rangle = -n_i | \phi_0 \rangle + \sqrt{n_i} | \phi_i \rangle = (-n_i \mathbf{1} + \mathbf{A}_i) | x \rangle \tag{4-15}$$

where $\vec{\sigma}_k$ is the pauli matrices associated with *k*-th vertex and $\pi_{kl}$ is permutation operator that swaps the state of *k*-th and lth qubits. Since the above relation holds for all x we have $\mathbf{A}_i = n_i \mathbf{1} + \frac{1}{2} \sum_{k,l: \partial(k,l)=i} \vec{\sigma}_k . \vec{\sigma}_l$. Inserting this relation into (4-14) we achieve

$$\mathbf{H} = \frac{1}{2t^*} \sum_{i=0}^{d} J_i \sum_{k,l: \partial(k,l)=i} \vec{\sigma}_k . \vec{\sigma}_l + \left( \sum_{i=0}^{d} J_i n_i / t \right) \mathbf{1} \tag{4-16}$$

The latter term is unity matrix so it adds only an overall phase to the quantum state; hence the dynamics of the system as well as the generated entanglement is insensitive to this part. Thus we can use

$$\mathbf{H} = \frac{1}{2t^*} \sum_{i=0}^{d} J_i \sum_{k,l: \partial(k,l)=i} \vec{\sigma}_k . \vec{\sigma}_l \tag{4-17}$$

We have shown that the proper Hamiltonian that enables creation of a Bell state between antipodal vertices in association scheme graphs is Heisenberg interaction between vertex pairs. Revealing such well known form of Hamiltonian enhances the physical motivation of the proposed systems. For more details about association scheme graphs see [12][13].

## 5. Examples

Here we are going to introduce some examples of "*I.S.G's*" and investigate the possibility of complete localization of quantum state on reference site and another stratum.

### Example 1

Consider the graph depicted in Fig 1

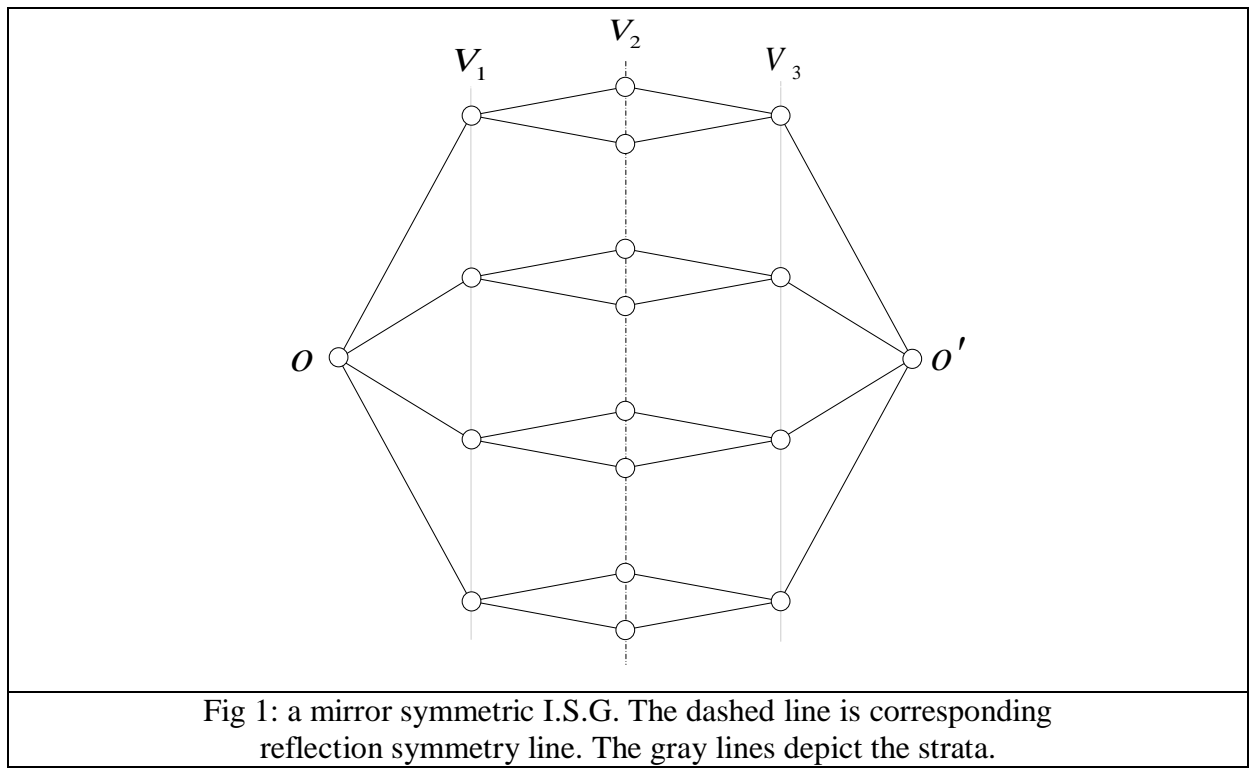

Fig 1: a mirror symmetric I.S.G. The dashed line is corresponding reflection symmetry line. The gray lines depict the strata.

This graph is a mirror symmetric I.S.G and due to the theorem 1 its Szego-Jacobi parameters are $\omega_1 = \omega_4 = 4$, $\omega_2 = \omega_3 = 2$, $\alpha_i = 0$. The matrix $P$ is obtained as

$$P = \begin{pmatrix} 1 & 1 & 1 & 1 & 1 \\ 0 & -1 & 1 & 1.4142 & -1.4142 \\ -1.4142 & 0 & 0 & 1.4142 & 1.4142 \\ 0 & 1 & -1 & 1.4142 & -1.4142 \\ 1 & -1 & -1 & 1 & 1 \end{pmatrix}$$

We see that only the last row of $P$ obeys the constraint (3-9) that corresponds to the antipode vertex of the reference site. These values are in accordance with the results of subsection 4-1 in the possibility of Bell state generation between antipodes of symmetric I.S.G's.

## Example 2

Consider the Tchebichef graph of the first kind [8][9] that is depicted in Fig 2

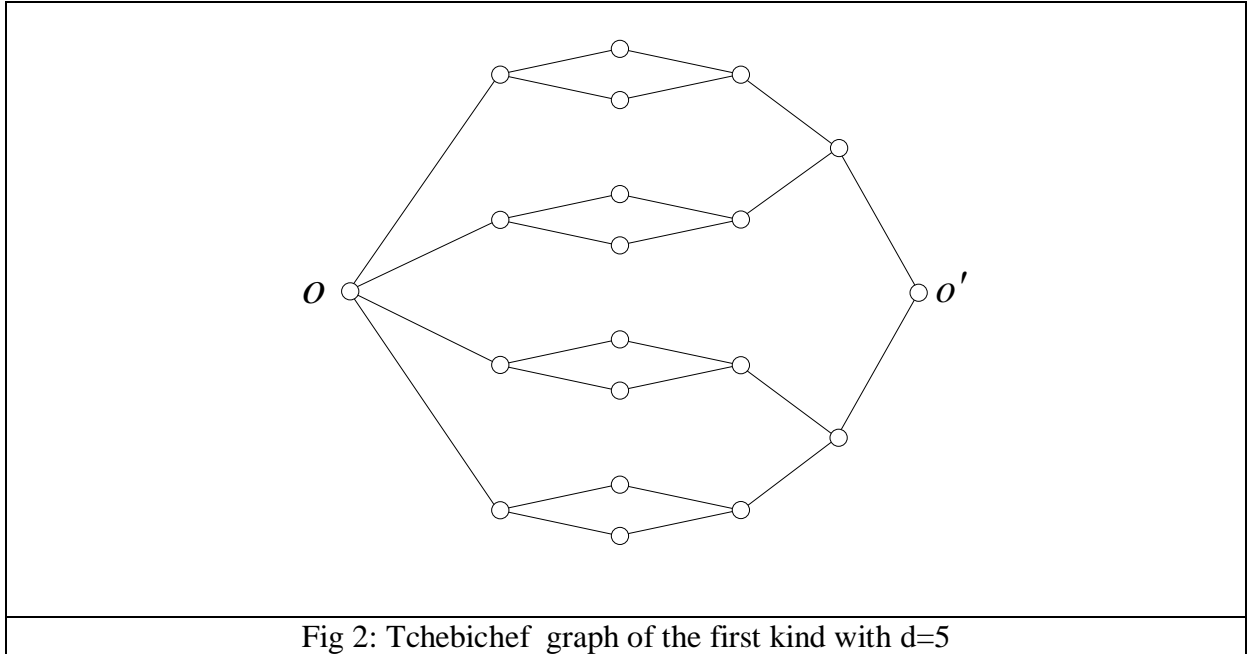

Fig 2: Tchebichef graph of the first kind with d=5

Due to theorem 1 the Szego-Jacobi parameters are $\omega_1 = 4$, $\omega_2 = \omega_3 = \omega_4 = \omega_5 = 2$, $\alpha_i = 0$. The matrix $P$ is obtained as

$$P = \begin{pmatrix} 1 & 1 & 1 & 1 & 1 & 1 \\ -1 & 1 & \frac{\sqrt{3}+1}{2} & \frac{1-\sqrt{3}}{2} & \frac{\sqrt{3}-1}{2} & -\frac{\sqrt{3}+1}{2} \\ 0 & 0 & \frac{\sqrt{6}}{2} & -\frac{\sqrt{6}}{2} & -\frac{\sqrt{6}}{2} & \frac{\sqrt{6}}{2} \\ 1 & -1 & 1 & 1 & -1 & -1 \\ -\sqrt{2} & -\sqrt{2} & \frac{\sqrt{2}}{2} & \frac{\sqrt{2}}{2} & \frac{\sqrt{2}}{2} & \frac{\sqrt{2}}{2} \\ 1 & -1 & \frac{\sqrt{3}-1}{2} & -\frac{\sqrt{3}+1}{2} & \frac{\sqrt{3}+1}{2} & \frac{1-\sqrt{3}}{2} \end{pmatrix}$$

We see that the fourth and fifth row of $P$ obey the constraint (3-9). This means that the quantum state can be completely localized on the reference site and one of the thirst or fourth stratum, but since these strata aren't single element, the Bell state can not be generated between vertices. This example shows that for complete localization of quantum state the elements of corresponding row of $P$ aren't necessarily +1 or -1. This example also shows that in some of antipodal I.S.G's the Bell state can not be generated between antipodes.

### Examples of antipodal association scheme graphs

In this part we introduce some antipodal association scheme graph and also present their corresponding Szego-Jacobi coefficients. For more details about these graphs refer to the corresponding references.

**Cycle graph** $C_{2m}$

The Szego-Jacobi coefficients of this network are given by $\omega_1 = \omega_m = 2$, $\omega_2 = \omega_3 = \ldots = \omega_{m-1} = 1$, $\alpha_i = 0$ [13]

**Hypercube**

The Szego-Jacobi coefficients of a Hypercube of dimension $d$ are given by $\omega_i = i(d-i+1)$, $\alpha_i = 0$ [13]

**Hadamard networks**

The Szego-Jacobi coefficients of this network are given by $\omega_1 = \omega_4 = 12$, $\omega_2 = \omega_3 = 66$, $\alpha_i = 0$ [14]

**Doubled odd graph DO(4)**

The Szego-Jacobi coefficients of this network are given by $\omega_1 = \omega_4 = 16$, $\omega_2 = \omega_3 = 36$, $\alpha_0 = \alpha_4 = 0$, $\alpha_1 = \alpha_3 = 6$, $\alpha_2 = 8$ [14]

**Johnson graph J(8,4)**

The Szego-Jacobi coefficients of this network are given by $\omega_1 = \omega_4 = \omega_7 = 4$, $\omega_2 = \omega_6 = 3$, $\omega_3 = \omega_5 = 6$, $\alpha_i = 0$ [14]

**Wells**

The Szego-Jacobi coefficients of this network are given by $\omega_1 = \omega_4 = 5$, $\omega_2 = \omega_3 = 4$, $\alpha_0 = \alpha_1 = \alpha_3 = \alpha_4 = 0$, $\alpha_2 = 3$ [15]

### 6. Summary

In this paper we attempted to generate a Bell state between distant vertices in permanently coupled quantum spin network interacting via invariant stratification graphs (I.S.G's). At the first step we established an upper bound over achievable entanglement between the reference site and the other vertices. Due to this upper bound we find that, creation of Bell state between reference site and a vertex is possible if the stratum of that vertex is single-element, e.g. antipodal I.S.G's. On the other hand the relevant upper bound of entanglement of a vertex can be saturated if the evolved quantum state of the system is completely localized over the reference site and the stratum of that vertex. This constraint is met if and only if the corresponding row of characteristic matrix **P(A)** is composed of at most two distinct values which are negatively inverse of each other. We introduce a special class of antipodal I.S.G's, which is called reflective, in which the antipode vertex obeys the characteristic constraint. This means that Bell state can be generated between the reference site and the farthest vertex from it. We also proved that association scheme graphs are I.S.G's. We also showed that the antipodal association scheme graphs are reflective so Bell state can be generated between the antipodal vertices. Moreover we see that the proper Hamiltonian that enables creation of Bell state between antipodal vertices in such graphs is Heisenberg interaction between vertex pairs. Revealing such well known form of Hamiltonian boosts the physical interests of the proposed systems. Recently, I.S.G's have also been considered for their capability for perfect state transfer [8]. So the present work enhances their priorities in quantum communication tasks.

**Acknowledgement**: M. Ghojavand acknowledges H. R. Afshar for careful reading of manuscript and his comments.